\documentclass[12pt]{iopart}

\usepackage{iopams,graphics,epsfig}  
\newcommand{\pr}{\mathop{\mathrm{pr}}\nolimits} 
\newcommand{\Sim}{\mathop{\mathrm{Sim}}}
\newcommand{\SL}{\mathop{\mathrm{SL}}}
\newcommand{\slmot}{\mathop{\mathrm{sl}}}
\newcommand{\GL}{\mathop{\mathrm{gl}}}
\newcommand{\Gell}{\mathop{\mathrm{GL}}}
\newcommand{\diff}{\mathop{\mathrm{diff}}}
 
\newcommand{\const}{\mathop{\mathrm{const.}}}

\begin{document}

\title{Difference schemes with point symmetries and their numerical tests}

\author{A~Bourlioux$^1$, C Cyr-Gagnon$^{1,3}$, and P Winternitz$^{1,2}$}
\address{$^1$ \ D\'epartement de math\'ematiques et de statistique, Universit\'e de Montr\'eal, C.P.~6128, succ. Centre-ville, Montr\'eal, Qu\'ebec H3C 3J7, Canada. \\
\ead{bourliou@dms.UMontreal.CA}}
\address{$^2$ \ Centre de recherches math\'ematiques, Universit\'e de Montr\'eal, C.P.~6128, succ. Centre-ville, Montr\'eal, Qu\'ebec H3C 3J7, Canada. \\
\ead{wintern@CRM.UMontreal.CA}}
\address{$^3$\ Present address: Cegep de St-Laurent, St. Laurent, Qu\'ebec H4L 3X7, Canada.\\
\ead{ccyrgagnon@cegep-st-laurent.qc.ca}}

\begin{abstract}
Symmetry preserving difference schemes approximating second and third order ordinary differential equations are presented. They have the same three or four-dimensional symmetry groups as the original differential equations. The new difference schemes are tested as numerical methods. The obtained numerical solutions are shown to be much more accurate than those obtained by standard methods without an increase in cost. For an example involving a solution with a singularity in the integration region the symmetry
preserving scheme, contrary to standard ones, provides solutions valid beyond the singular point.
\end{abstract}



\maketitle

\section{Introduction} \label{AP:sec1}
The purpose of this article is to present some new difference schemes, having the same Lie point symmetry groups as the ordinary differential equations they approximate. We test these schemes as numerical methods and compare them with standard schemes.

This is part of a general program, the aim of which is to turn Lie group theory into an efficient tool for solving difference equations. Continuous symmetries of discrete equations have been intensively studied during the last 20 years or so.

For recent reviews containing extensive lists of references to the original papers, see \cite{AP:ref1, AP:ref2}. In this article we restrict ourselves to one specific aspect of this approach, the symmetry preserving discretization of ordinary difference equations and its applications in numerical analysis.

To present the basic ideas, let us first consider an ordinary differential equation (ODE)
\begin{equation} \label{AP:eq1.1}
E \equiv y^{(n)} - F(x, y, y', \dots , y^{(n-1)}) = 0.
\end{equation}
Its Lie point symmetry group $G$ consists of all local point transformations of the form
\begin{equation} \label{AP:eq1.2}
\tilde{x} = \Lambda_{\lambda}(x, y), \quad \tilde{y} = \Omega_{\lambda}(x, y)
\end{equation}
taking solutions $y(x)$ into solutions $\tilde{y}(\tilde{x})$ of the same equation ($\lambda$ represents group parameters). The Lie algebra $L$ of the symmetry group $G$ is realized by vector fields of the form
\begin{equation} \label{AP:eq1.3}
X = \xi(x, y)\partial x + \phi(x, y)\partial y.
\end{equation}
The algorithm for finding the symmetry algebra $L$ and the symmetry group $G$ for a given ODE (\ref{AP:eq1.1}) goes back to S.~Lie and is given in many books on the subject \cite{AP:ref3}. It consists of solving the determining equations resulting from the infinitesimal invariance requirement
\begin{equation} \label{AP:eq1.4}
\pr^{(n)} X(E)\big|_{E=0} = 0,
\end{equation}
where $\pr^{(n)}X$ is the $n$-th order prolongation of the vector field $X$ (acting on derivatives up to order $n$) \cite{AP:ref3}.

Let us now consider an ordinary difference scheme (O$\Delta$S), approximating the ODE (\ref{AP:eq1.1}). The scheme will consist of two equations relating the values of $(x, y)$ in $N$ different points, with $N \ge n + 1$
\begin{equation} \label{AP:eq1.5}
E_a(n, x_{n+K}, \dots , x_{n+L}, y_{n+K}, \dots y_{n+L}) = 0, \quad a \!=\! 1, 2, \ L \!- \!K \!=\! N\! -\! 1.
\end{equation}
In the continuous limit one equation, say $E_1$ goes into the ODE (\ref{AP:eq1.1}), the other reduces to an identity (like $0 = 0$). The two equations (\ref{AP:eq1.5}) should be such that if $N - 1$ values $(x_k, y_k)$ are given, we can calculate the $N$-th one. This is assured e.g.\ by imposing a condition on the Jacobian:
\begin{equation} \label{AP:eq1.6}
\frac{\partial(E_1, E_2)}{\partial(x_{n+L}, y_{n+L})} \ne 0.
\end{equation}

We wish to construct an O$\Delta$S that not only approximates eq.~(\ref{AP:eq1.1}), but has the same Lie point symmetry group $G$. This is achieved by constructing the scheme out of \emph{difference invariants} of the group $G$, or out of invariant manifolds. These are found using the vector fields (\ref{AP:eq1.3}), corresponding to the invariance algebra of the ODE (\ref{AP:eq1.1}). The vector fields are the same as in the continuous case, however they must be prolonged to all points of the lattice, involved in the system (\ref{AP:eq1.5}). We have
\begin{equation} \label{AP:eq1.7}
\pr X = \sum_{j=K}^L \{ \xi(x_{n+j}, y_{n+j}) \partial x_{n+j} + \phi(x_{n+j}, y_{n+j})\partial  y_{n+j}\}.
\end{equation}
The invariants satisfy
\begin{equation} \label{AP:eq1.8}
\pr X_a I(x_{n+j}, y_{n+j}) = 0, \quad a = 1, \dots M,
\end{equation}
where $\{X_1, \dots , X_M\}$ is a basis of the algebra $L$. The invariant manifolds satisfy the same equation, but only on a subspace where the matrix of coefficients
\[
\left(
\begin{array}{l}
\xi_{1,n+K},\dots , \xi_{1,n+L},  \phi_{1,n+K}, \dots , \phi_{1,n+L}\\
\vdots \\
\xi_{M,n+K},\dots , \xi_{M,n+L},  \phi_{M,n+K}, \dots , \phi_{M,n+L}
\end{array}
\right)
\]
is of lower rank.

The fact that difference schemes can be invariant under continuous Lie point transformations that act on the equations and on lattices was pointed out by Dorodnitsyn \cite{AP:ref2, AP:ref4}. This approach has been used to classify and solve three-point difference schemes \cite{AP:ref5, AP:ref6}. It has been shown that symmetry preserving discretizations of first order ODEs are exact, i.e.\ the solutions of the ODEs and the invariant O$\Delta$S coincide exactly \cite{AP:ref7}. The complementary problem, in which an O$\Delta$S is given and we wish to find its Lie point symmetries was solved in \cite{AP:ref8a}.

In Section~\ref{AP:sec2} we discretize a second order nonlinear ODE with a three-dimensional solvable symmetry algebra. Section~\ref{AP:sec3} is devoted to a discretization of several third order nonlinear ODEs with three, or four-dimensional solvable symmetry algebras. Equations invariant under the simple group $\SL(2, \mathbb{R})$, or the reductive one $\Gell(2, \mathbb{R})$ are discretized in Section~\ref{AP:sec4}. The obtained invariant difference schemes are tested in  Section~\ref{AP:sec5}. They are shown to be considerably more accurate then the corresponding standard schemes. Moreover, in the study of a singular solution the invariant schemes turn out to have a qualitative advantage: they make it possible to integrate numerically beyond the singularity. 

\section{Example 1 : A second order ODE invariant under a solvable Lie group}\label{AP:sec2}

Let us consider the second order ODE
\begin{equation} \label{AP:eq2.1}
x^2y'' + 4xy' + 2y = (2xy + x^2y')^{(k-2)/(k-1)}, \quad k \ne 0, {\textstyle\frac12}, 1, 2.
\end{equation}
Its symmetry algebra has a basis given by
\begin{equation} \label{AP:eq2.2}
X_1 = \frac{\partial}{\partial x} - \frac{2y}{x} \frac{\partial}{\partial y}, \quad
X_2 = \frac{1}{x^2}\frac{\partial}{\partial y}, \quad X_3 = x \frac{\partial}{\partial  x} + (k -  2)y\frac{\partial}{\partial y}
\end{equation}
(for $k = 0$, ${\textstyle{\frac12}}$ and $2$ the symmetry algebra is larger and the equation is linear, or linearizable). Equation~(\ref{AP:eq2.1}) could be simplified by a transformation taking the algebra (\ref{AP:eq2.2}) into its standard form, but we are interested in discretizing it without prior simplifications.

We note that the general solution of (\ref{AP:eq2.1}) is
\begin{equation} \label{AP:eq2.3}
y(x) = \left(\frac{1}{k-1}\right)^{k-1} \frac{1}{kx^2}(x - x_0) + \frac{y_0}{x}
\end{equation}
where $x_0$ and $y_0$ are integration constants.

Now let us derive an O$\Delta$S approximating the ODE (\ref{AP:eq2.1}), invariant under the Lie group generated by (\ref{AP:eq2.2}). We consider 3 points on a line $x_{n-1}$, $x_n$, $x_{n+1}$ and the corresponding values $y_k = y(x_k)$. The invariance condition
\begin{equation} \label{AP:eq2.4}
\pr XF(x_{n-1}, x_n, x_{n+1}, y_{n-1}, y_n, y_{n+1})=0
\end{equation}
with $\pr X$ as in eq.~(\ref{AP:eq1.7}) yields 3 elementary invariants:
\begin{equation} \label{AP:eq2.5}
\xi_1
= \frac{x_{n+1} - x_n}{x_n - x_{n-1}}, \quad
\xi_2
= \frac{x_{n+1}^2 y_{n+1}- x_n^2y_n}{(x_{n+1} - x_n)^k}, \quad 
\xi_3
= \frac{x_n^2y_n - x_{n-1}^2y_{n-1}}{(x_n - x_{n-1})^k}.
\end{equation}
We put $h_{n+1} = x_{n+1} - x_n$, $h_n = x_n - x_{n-1}$ and expand $y_{n \pm 1} = y(x_{n \pm 1})$ into Taylor series about $x = x_n$. We obtain
\begin{eqnarray} \label{AP:eq2.6}
&\frac{2\xi_1}{\xi_1+1}\biggl( \xi_2  - \frac{1}{(\xi_1)^{k-1}}\xi_3\biggr)  
 =(h_{n+1})^{2-k}  \{ (x^2y'' + 4xy' + 2y)\nonumber \\
&\hspace{2in}  + {\textstyle\frac{1}{3}} (h_{n+1} - h_n)(x^2y''' + 6xy'' + 6y') + 0(\varepsilon^2) \}\nonumber \\ 
&\frac{1}{2} \biggl[ \xi_2 + \frac{1}{(\xi_1)^{k-1}}\xi_3\biggr]^{(k-2)/(k-1)} = (h_{n+1})^{2-k} (x^2y' + 2xy)^{(k-2)/(k-1)} \label{AP:eq2.7}\\
&\hspace{1.3in}\times \Biggl\{ 1 + \frac{k-2}{k-1}(h_{n+1} - h_n) \frac{x^2y'' + 4xy' + 2y}{x^2y' + 2xy} + O(\varepsilon^2)\Biggr\}\nonumber
\end{eqnarray}
where we assume that $h_{n+1}$ and $h_n$ are of order $\varepsilon$.

We see that the two equations
\begin{eqnarray} 
&\frac{2\xi_1}{\xi_1+1} \Biggl( \xi_2 - \frac{1}{(\xi_1)^{k-1}}\xi_3 \Biggr) = \frac{1}{2} 
\Biggl[ \xi_2 + \frac{1}{(\xi_1)^{k-1}}\xi_3 \Biggr]^{(k-2)/(k-1)},
\label{AP:eq2.8}\\
& {\xi_1}= K,\label{AP:eq2.9}
\end{eqnarray}
with $K = \const$ provide an invariant O$\Delta$S approximating eq.~(\ref{AP:eq2.1}). In general this is a first order approximation (of order $\varepsilon$). If we choose $K = 1$ in (\ref{AP:eq2.9}), the first order terms drop out and we obtain a second order scheme (and a uniform lattice).

\section{Examples of third order ODEs invariant under solvable Lie groups}\label{AP:sec3}

\subsection{General comments}\label{AP:subsec3.1}
A third order ODE can have a Lie point symmetry group of dimension $\dim \mathcal{L} = N$, $0 \le N \le 7$. The maximal dimension $N = 7$ occurs only for linear equations that can be transformed into ${y'''} = 0$ by a point transformation \cite{AP:ref3, AP:ref9}. We shall consider examples of equations with $N = 3$ and $N = 4$.

In order to approximate a third order ODE we must consider at least 4 points in a stencil. We denote the points
\begin{equation} \label{AP:eq3.1}
(x_{n+k}, y_{n+k}), \quad -1 \le k \le 2,
\end{equation}
and put
\begin{equation} \label{AP:eq3.2}
h_{n+2} = x_{n+2} - x_{n+1}, \quad h_{n+1} = x_{n+1} - x_n, \quad h_n = x_n - x_{n-1}.
\end{equation}
In the continuous limit we put
\begin{equation} \label{AP:eq3.3}
h_{n+j} = \alpha_j \varepsilon, \quad j = 0, 1, 2
\end{equation}
where $\alpha_j$ are constants of the order of $1$ (not necessarily all equal).

\subsection{Example 2 : An {\rm{ODE}} invariant under the similitude group of a Euclidean plane}\label{AP:subsec3.2}
Let us consider the four-dimensional Lie algebra
\begin{equation} \label{AP:eq3.4}
X_1 = \frac{\partial}{\partial x}, \quad 
X_2 = \frac{\partial}{\partial y}, \quad
X_3 = y \frac{\partial}{\partial x} - x\frac{\partial}{\partial y}, \quad
X_4 = x \frac{\partial}{\partial x} + y\frac{\partial}{\partial y}, 
\end{equation}
generating translations, rotations and dilations in the $(x, y)$ plane, respectively. The Euclidean algebra $\{X_1, X_2, X_3 \}$ allows two independent differential invariants in the space $\{ x, y, y', y'', y'''\}$
\begin{equation} \label{AP:eq3.5} 
I_1 = \frac{y''}{(1 + y^{\prime 2})^{3/2}}, \quad
I_2 = \frac{(1+y^{'2}) y''' - 3y'y^{''2}}{(1 + y^{'2})^3}
\end{equation}
and the invariant ODE is
\begin{equation} \label{AP:eq3.6} 
I_2 = F(I_1),
\end{equation}
where $F(z)$ is an arbitrary function.

Invariance under dilations corresponding to $X_4$ implies $F(z) = Kz^2$ and the invariant ODE is
\begin{equation} \label{AP:eq3.7} 
(1 + y^{'2})y''' - 3y'y^{''2} = Ky^{''2}
\end{equation}
where $K$ is a constant. The general solution of eq.~(\ref{AP:eq3.7}) can be given in implicit form as
\begin{equation} \label{AP:eq3.8}
y(x) = \int_0^x u(t)\, dt + C_3, \quad x = C_1 \int_0^u \frac{e^{-K \arctan s}}{(1 + s^2)^{3/2}} \, ds + C_2,
\end{equation}
where $C_1$, $C_2$, and $C_3$ are constants.

The Euclidean Lie group corresponding to $\{X_1, X_2, X_3\}$ allows 5 functionally independent difference invariants in the space with local coordinates (\ref{AP:eq3.1}). We choose the following basis for the invariants:
\begin{eqnarray}\label{AP:eq3.9}
\xi_1 
&= h_{n+2} \Bigg[ 1 + \bigg(\frac{y_{n+2}-y_{n+1}}{h_{n+2}}\Bigg)^2\Bigg]^{1/2},\nonumber\\
\xi_2 
&= h_{n+1} \Bigg[ 1 + \bigg(\frac{y_{n+1}-y_n}{h_{n+1}}\Bigg)^2\Bigg]^{1/2},\nonumber\\
\xi_3 
&= h_n \Bigg[ 1 + \bigg(\frac{y_n-y_{n-1}}{h_n}\Bigg)^2\Bigg]^{1/2},\\
\xi_4
&= (y_{n+2} - y_{n+1})h_{n+1} - (y_{n+1} - y_n)h_{n+2},\nonumber \\
\xi_5
&= (y_{n+1} - y_n)h_n - (y_n - y_{n-1})h_{n+1}. \nonumber 
\end{eqnarray}
From these we can form invariants that approximate the differential invariants $I_1$ and $I_2$ of eq.~(\ref{AP:eq3.5}). To see this we expand $y_{n+2} = y(x_n + h_{n+1} + h_{n+2})$, $y_{n+1} = y(x_n + h_{n+1})$ and $y_{n-1} = y(x_n - h_n)$ into Taylor series about $x_n$ and obtain
\begin{eqnarray}\label{AP:eq3.10}
J_2 
&= \frac{6}{\xi_1 + \xi_2 + \xi_3} \Biggl( \frac{\xi_4}{\xi_1\xi_2(\xi_1 + \xi_2)} - \frac{\xi_5}{\xi_2\xi_3(\xi_2 + \xi_3)} \Bigg) \\
&\quad = \frac{1}{(1 + y^{'2})^3} \Biggl\{ \Biggl[ (1 + y^{'2})y''' - 3y' y^{''2}\Biggr] \nonumber\\
&\quad \quad \quad \quad  + \Biggl( \frac{h_{n+2} + 2h_{n+1}-h_n}{4} \Biggr)
\Biggl[ (1 + y^{'2})y^{iv} - 10 y'y''y''' + 15 y^{'2}y^{''3}\Biggr]\nonumber\\
&\quad \quad \quad \quad \quad  - \frac{3}{8} \frac{y^{'''3}}{1 + y^{'2}} \frac{2h^2_{n+2} + 7h_{n+2}h_{n+1} + 4h_{n+1}^2 + h_{n+1} h_n-2h_n^2}{(h_{n+2} + h_{n+1} + h_n)}\Biggr\}\nonumber
\end{eqnarray}

\begin{eqnarray}\label{AP:eq3.11}
J_1
&= \frac{2\alpha \xi_4}{\xi_1\xi_2(\xi_1 + \xi_2)} + \frac{2\beta \xi_5}{(\xi_2 \xi_3)(\xi_2 + \xi_3)}\\
&= \frac{1}{(1 + y^{'2})^{3/2}} \Biggl\{ y'' + \frac{1}{3(1+y^{'2})} [ (1+ y^{'2})y''' - 3y'y^{''2}] \nonumber\\
&\qquad\qquad\qquad\qquad\qquad \times
[\alpha (h_{n+2} + 2h_{n+1}) + \beta(h_{n+1} - h_n]\Biggr\}, \nonumber\\
\alpha &+ \beta   = 1. \nonumber
\end{eqnarray}
An invariant O$\Delta$S approximating eq.~(\ref{AP:eq3.6}) is given by
\begin{equation}\label{AP:eq3.12}
J_2 = F(J_1)
\end{equation}
on the lattice:
\begin{equation}\label{AP:eq3.13}
A\xi_1 + B\xi_2 + C\xi_3 = 0,
\end{equation}
where $A$, $B$ and $C$ are constants.

In particular the ODE (\ref{AP:eq3.7}) invariant under the similitude group $\Sim(2)$ is approximated by
\begin{equation}\label{AP:eq3.14}
J_2 = KJ_1^2
\end{equation}
on the lattice (\ref{AP:eq3.13}) which is also invariant under $\Sim(2)$.

Other invariant lattices can be formed out of the invariants (\ref{AP:eq3.9}), for instance in Subection~\ref{AP:subsec5.3} we choose
\begin{equation}\label{AP:eq3.15}
\frac{\xi_1}{\xi_2} = \frac{\xi_2}{\xi_3}.
\end{equation}

Generally speaking (\ref{AP:eq3.13}) and (\ref{AP:eq3.14}) (or (\ref{AP:eq3.15})) provide a first order approximation (of order $\varepsilon$ if $h_{n+2}$, $h_{n+1}$ and $h_n$ are of order $\varepsilon$). For a special value of $K$ we can cancel first order terms in $\varepsilon$ and obtain a second order approximation, namely
\[
K = \frac{\sqrt{3}}{2}, \quad \alpha = \beta = {\textstyle\frac{1}{2}}, \quad C = -A, \quad B = 2A.
\]

\subsection{Example 3 : Equations invariant under a Euclidean Lie group}\label{AP:subsec3.3}
Let us consider a different realization of the Euclidean and similitude Lie algebras, namely
\begin{equation}\label{AP:eq3.16}
X_1 = \frac{\partial}{\partial y}, \quad
X_2 = x\frac{\partial}{\partial y}, \quad
X_3 = (1 + x^2)\frac{\partial}{\partial x} + xy \frac{\partial}{\partial y}, \quad
X_4 = y \frac{\partial}{\partial y}.
\end{equation}
This algebra is isomorphic to (\ref{AP:eq3.4}) but cannot be transformed into it by a transformation of variables. The Euclidean Lie group corresponding to $\{ X_1, X_2, X_3\}$ allows two independent differential invariants of order 3 or less. We choose them to be
\begin{eqnarray}\label{AP:eq3.17}
I_1 &= (1 + x^2)^{3/2} y'', \\
I_2 &= [(1 + x^2)y''' + 3xy''](1 + x^2)^{3/2}.\nonumber
\end{eqnarray}
The invariant third order ODE is
\begin{equation}\label{AP:eq3.18}
I_2 = F(I_1)
\end{equation}
where $F(z)$ is an arbitrary function.  If we also require invariance under the dilations generated by $X_4$, we obtain $F(z) = Az$ and the equation is a linear one.

The five functionally independent difference invariants in the space (\ref{AP:eq3.1}) allowed by the Euclidean group generated by $\{ X_1, X_2, X_3\}$ are
\begin{eqnarray}\label{AP:eq3.19}
\xi_1 
&= (1 + x_n^2)^{1/2} \Biggl[ \frac{y_{n+1} - y_n}{x_{n+1} - x_n} - \frac{y_n-y_{n-1}}{x_n-x_{n-1}} \Biggr], \nonumber \\
\xi_2 
&= (1 + x_{n+1}^2)^{1/2} \Biggl[ \frac{y_{n+2} - y_{n+1}}{x_{n+2} - x_{n+1}} - \frac{y_{n+1}-y_n}{x_{n+1}-x_n} \Biggr], \\
\xi_3
&= \frac{x_n - x_{n-1}}{1 + x_nx_{n-1}}, \quad
\xi_4 = \frac{x_{n+1} - x_n}{1 + x_n x_{n+1}}, \quad
\xi_5 = \frac{x_{n+2} - x_{n+1}}{1 + x_{n+1}x_{n+2}}. \nonumber
\end{eqnarray}
Expanding into Taylor series about the point $x \equiv x_n$ we find
\begin{eqnarray}\label{AP:eq3.20}
\frac{2\xi_1}{\xi_3 \!+\! \xi_4}
&\!= \!(1 \!+\! x^2)^{3/2} \Biggl[ y'' \!+\! \frac{h_{n+1} \!-\! h_n}{3}\Biggl( y''' \!+\! \frac{3x_n}{1 \!+\! x_n^2} y''\Biggr) \!+\! O(\varepsilon^2) \Biggr],\\
\frac{2\xi_2}{\xi_4 \!+\! \xi_5}
&= (1 \!+\! x^2)^{3/2} \Biggl[ y'' \!+\! \frac{h_{n+2} \!+\! 2h_{n+1}}{3}\Biggl( y''' \!+\! \frac{3x_n}{1 \!+\! x_n^2} y''\Biggr) \!+\! O(\varepsilon^2) \Biggr].\nonumber
\end{eqnarray}
We have assumed that $h_n$, $h_{n+1}$ and $h_{n+2}$ are all of order $\varepsilon$ (but not necessarily equal).

From eq.~(\ref{AP:eq3.20}) we obtain
\begin{eqnarray}\label{AP:eq3.21}
J_1
&= \frac{2\alpha \xi_1}{\xi_3 + \xi_4} + \frac{2\beta \xi_2}{\xi_4 + \xi_5} = (1 +x^2)^{3/2} y'' + O(\varepsilon), \quad \alpha + \beta =1, \\
J_2
&= \frac{6}{\xi_3 + \xi_4 +\xi_5}  \Biggl(\frac{\xi_2}{\xi_4 + \xi_5} - \frac{\xi_1}{\xi_3 +\xi_4} \Biggr) \nonumber \\
	&\qquad \qquad \qquad \qquad \qquad \qquad = (1 +x^2)^{3/2}[(1+x^2)y''' + 3xy'']  + O(\varepsilon). \nonumber
\end{eqnarray}
Thus an invariant 0$\Delta$S approximating eq.~(\ref{AP:eq3.18}) is given by
\begin{equation}\label{AP:eq3.22}
J_2 = F(J_1), \quad a\xi_3 + b\xi_4 + c\xi_5 = 0
\end{equation}
where $a$, $b$ and $c$ are constants. In general this will be a first order scheme. For certain functions $F$ the scheme can be improved to a second order one by an appropriate choice of the constants $\alpha$, $\beta$, $a$, $b$ and $c$. We shall not go into that here.

\section{\mathversion{bold}Third order equations invariant under $\SL(2, \mathbb{R})$}\label{AP:sec4}

Four inequivalent realizations of $\slmot(2, \mathbb{R})$ as subalgebras of $\diff (2, \mathbb{R})$ exist \cite{AP:ref5, AP:ref10}. We shall consider two of them here.

\subsection{Example 4 : First $\slmot(2, \mathbb{R})$ algebra}

The first $\slmot(2, \mathbb{R})$ algebra $S_1$ is given by
\begin{equation}\label{AP:eq4.1}
X_1 = \frac{\partial}{\partial y}, \quad 
X_2 = x \frac{\partial}{\partial x} + y \frac{\partial}{\partial y}, \quad
X_3 = 2xy \frac{\partial}{\partial x}+  y^2 \frac{\partial}{\partial y}.
\end{equation}
It allows one second order and one third order differential invariant:
\begin{equation}\label{AP:eq4.2}
I_1 = \frac{2xy'' + y'}{y{'3}}, \quad I_2 = \frac{x^2(y' y''' - 3y^{''2})}{y^{'5}}.
\end{equation}
The most general invariant third order ODE is hence
\begin{equation}\label{AP:eq4.3}
I_2 = F(I_1), 
\end{equation}
where $F(z)$ is an arbitrary function. The algebra (\ref{AP:eq4.1}) can be extended to $\GL(2, \mathbb{R})$ by adding the operator 
\begin{equation}\label{AP:eq4.4}
X_4 = x\frac{\partial}{\partial x}.
\end{equation}
Requiring invariance under the corresponding $\Gell(2, \mathbb{R})$ group restricts $F(z)$ to $F(z) = Az^{3/2}$ and we obtain the ODE
\begin{equation}\label{AP:eq4.5}
x^2(y'y''' - 3y^{''2}) = Ay^{' 1/2} (2xy'' + y')^{3/2}.
\end{equation}
Five independent $\SL(2, \mathbb{R})$ difference invariants are
\begin{eqnarray} \label{AP:eq4.6}
&\xi_1 = \frac{1}{\sqrt{x_{n+1}x_{n+2}}} (y_{n+2} - y_{n+1}), &\quad
\xi_2 = \frac{1}{\sqrt{x_n x_{n+1}}}(y_{n+1} - y_n), \nonumber\\
&\xi_3 = \frac{1}{\sqrt{x_{n-1}x_n}} (y_n - y_{n-1}), &\quad
\xi_4 = \frac{1}{\sqrt{x_n x_{n+2}}}(y_{n+2} - y_n), \\
&\xi_5 = \frac{1}{\sqrt{x_{n+1}x_{n-1}}} (y_{n+1} - y_{n-1}). & \nonumber
\end{eqnarray}
From these we form
\begin{eqnarray} 
J_2
&= 12 \frac{(\xi_4 - \xi_1 - \xi_2)(\xi_2 + \xi_3)\xi_3 - (\xi_5 - \xi_2 - \xi_3)\xi_1(\xi_1 + \xi_2)}{\xi_1\xi_2\xi_3(\xi_1 + \xi_2)(\xi_2 + \xi_3)(\xi_1 + \xi_2 + \xi_3)} \label{AP:eq4.7}\\
&= \frac{x^2(y'y''' - 3y^{''2})}{y^{'5}} + \varepsilon \phi, \nonumber\\
J_1 
&= 8 \Biggl[ \alpha \frac{\xi_4 - \xi_1 - \xi_2}{\xi_1\xi_2(\xi_1 + \xi_2)} + (1-\alpha)  \frac{\xi_5 - \xi_2 - \xi_3}{\xi_2\xi_3(\xi_2 + \xi_3)}\Biggr] \label{AP:eq4.8}\\
&= \frac{2 xy'' + y'}{y^{'3}} + \frac{2}{3} 
[ \alpha (h_{++} + 2h_+) + (1-\alpha)(h_+ - h)] \frac{x(y'y''' - 3y^{''2}}{y^{'4}}\nonumber \\
&\qquad\qquad \qquad \qquad \qquad \qquad\qquad \qquad \qquad\qquad \qquad \qquad     + \varepsilon^{2}\psi. \nonumber
\end{eqnarray}
where $\phi$ and $\psi$ are some functions of $x$, $y'$, $y''$, $y'''$ and $y^{iv}$ and $0 \le \alpha \le 1$ is a constant.

An invariant scheme approximating eq.~(\ref{AP:eq4.3}) is given by
\begin{equation} \label{AP:eq4.9}
J_2 = F(J_1),
\end{equation}
\begin{equation}\label{AP:eq4.10}
A\xi_1 + B\xi_2 + C\xi_3 + D\xi_4 + E\xi_5 =0, 
\end{equation}
where $A, \dots , E$ are constants. To lowest orders (\ref{AP:eq4.10}) yields:
\begin{eqnarray} \label{AP:eq4.11}
&\{ Ah_{n+2} + Bh_{n+1} +Ch_n + D(h_{n+1} + h_{n+2}) + E(h_{n+1} + h_n)\}
\frac{y'}{x} \\
&\qquad + \{ A(h_{n+2} + 2h_{n+1})h_{n+1} + Bh_n^2 - Ch_{n-1}^2 + D(h_{n+1} + h_{n+2})^2 \nonumber\\
&\qquad\qquad\qquad\qquad\qquad \qquad + E(h_{n+1}^2 - h_n^2)\} \frac{xy'' - y'}{2x^2} = 0. \nonumber
\end{eqnarray}
In general, the scheme is a first order one, i.e.\ all $h_j$ go to zero like $h_j = a_j \varepsilon$, then the error in (\ref{AP:eq4.9}), (\ref{AP:eq4.10}) goes to zero like $\varepsilon^1$. For specific functions $F(z)$ the accuracy can be improved by an appropriate choice of the constants $\alpha$ and $A, \dots , E$.

\subsection{Example 5 : Second $\slmot(2, \mathbb{R})$ algebra}
The second $\slmot(2, \mathbb{R})$ algebra $S_2$ has a basis given by
\begin{equation} \label{AP:eq4.12}
X_1 = \frac{\partial}{\partial y}, \quad X_2 = y \frac{\partial}{\partial y}, \quad X_3 = y^2 \frac{\partial}{\partial y}.
\end{equation}
It can be embedded into the algebra $\slmot(2, \mathbb{R}) \oplus \slmot(2, \mathbb{R})$ by adding
\begin{equation} \label{AP:eq4.13}
X_4 = \frac{\partial}{\partial x}, \quad X_5 = x \frac{\partial}{\partial x}, \quad X_6 = x^2 \frac{\partial}{\partial x}.
\end{equation}
The Lie group $\SL(2, \mathbb{R})$ generated by $S_2$ has two differential invariants in the considered space, namely
\begin{equation} \label{AP:eq4.14}
I_1 = \frac{1}{y^{'2}} \Biggr(y'y''' - \frac{3}{2} y^{''2}\Biggr), \quad I_2 = x.
\end{equation}
The invariant ODE is
\begin{equation} \label{AP:eq4.15}
\frac{1}{y^{'2}}\Biggl( y'y''' - \frac{3}{2} y^{''2}\Biggr) = F(x).
\end{equation}
Requiring invariance under the $\Gell(2, \mathbb{R})$ group that includes $X_4$ in its Lie algebra reduces (\ref{AP:eq4.15}) to\
\begin{equation} \label{AP:eq4.16}
\frac{1}{y^{'2}}\Biggl( y'y''' - \frac{3}{2} y^{''2}\Biggr) = K
\end{equation}
where $K$ is a constant.  

A larger invariance group is obtained only for $K = 0$. In this case the equation is invariant under $\SL(2, \mathbb{R}) \otimes \SL(2, \mathbb{R})$, generated by (\ref{AP:eq4.12}) and  (\ref{AP:eq4.13}).

The difference invariants corresponding to the algebra (\ref{AP:eq4.12}) are
\begin{equation} \label{AP:eq4.17}
R = \frac{(y_{n+2} - y_n)(y_{n+1} - y_{n-1})}{(y_{n+2} - y_{n+1})(y_n - y_{n-1})}, \quad x, h_{n+2}, h_{n+1}, h_n.
\end{equation}
We have
\begin{eqnarray} \label{AP:eq4.18}
J_1
&= \frac{6h_{n+2} h_n}{h_{n+1}(h_{n+1}+h_{n+2})(h_n + h_{n+1})(h_{n+2} +h_{n+1} +h_n) } \\
&\qquad\qquad\qquad\qquad\qquad\qquad\qquad  \times \Biggl[ \frac{(h_{n+2} + h_{n+1})(h_{n+1} + h_n)}{h_n h_{n+1}} - R \Biggr] \nonumber\\
&= \frac{1}{y^{'2}} \Biggl[ y'y''' - \frac{3}{2} y^{''2}\Biggr] + O(\varepsilon). \nonumber
\end{eqnarray}
An invariant O$\Delta$S approximating eq.~(\ref{AP:eq4.15}) is
\begin{equation} \label{AP:eq4.19}
J_1 = F(x_n, h_n, h_{n+1}, h_{n+2}), \quad 
\phi(x_n, h_n, h_{n+1}, h_{n+2}) = 0
\end{equation}
with
\begin{eqnarray} 
F(x_n, 0, 0, 0) &= F(x), \label{AP:eq4.20}\\
\phi(x_n, 0, 0, 0) &=0. \label{AP:eq4.21}
\end{eqnarray}

If we require invariance under the group corresponding to $\{ X_1, X_2, X_3, X_4 \}$ we must take $F(x) = K$ and the lattice will depend only on $h_{n+2}$, $h_{n+1}$, and $h_n$. For instance we can take the lattice to be given by
\begin{equation} \label{AP:eq4.22}
\alpha h_{n+2} + \beta h_{n+1} + \gamma h_n = 0
\end{equation}
and the constants $\alpha$, $\beta$, $\gamma$ can be chosen to improve the approximation.

An O$\Delta$S invariant under $\SL(2, \mathbb{R}) \otimes \SL(2, \mathbb{R})$ that approximates eq.~(\ref{AP:eq4.16}) for $K = 0$ is
\begin{equation}
\frac{(x_{n+2} - x_n)(x_{n+1} - x_{n-1})}{(x_{n+2} - x_{n+1})(x_n - x_{n-1})} - 
\frac{(y_{n+2} - y_n)(y_{n+1} - y_{n-1})}
{(y_{n+2} - y_{n+1})(y_n - y_{n-1})} = 0,\label{AP:eq4.23}
\end{equation}
\begin{equation}
\frac{(x_{n+2} - x_n)}{(x_{n+2} - x_{n+1})} \frac{(x_{n+1} - x_{n-1})}{(x_n - x_{n-1})} = K_0. \label{AP:eq4.24}
\end{equation}
For $K_0 = 4$ this scheme is an exact one. Indeed, the equation
\begin{equation} \label{AP:eq4.25}
y'y''' - {\textstyle\frac{3}{2}} y^{''2} = 0
\end{equation}
has two families of solutions
\begin{equation} \label{AP:eq4.26}
y = \frac{1}{ax+b} + c \quad {\rm and}\quad y = \alpha x + \beta
\end{equation}
where $a$, $b$, $c$, $\alpha$ and $\beta$ are integration constants eq.~(\ref{AP:eq4.24}) with $K_0 = 0$ has two families of solutions
\begin{equation} \label{AP:eq4.27}
x_n = \frac{1}{an+b} +c, \quad x_n = \alpha n +\beta.
\end{equation}
On the lattice (\ref{AP:eq4.27}) the functions (\ref{AP:eq4.26}) solve (\ref{AP:eq4.23}) exactly.

In this example the underlying Lie group $\SL(2, \mathbb{R})$ plays a specially prominent role. The group is the group of projective transformations of the real line (the variable $y$). Its invariant $I_1$ is the Schwarzian derivative of the variable $y$. Projective transformations can be used to transform any three points on the projective line into any other three chosen points. Given four points, e.g. $y_{n-1}, y_n, y_{n+1}, y_{n+2}$ we can form precisely one projective invariant out of them, namely the anharmonic ratio $R$
of (\ref{AP:eq4.17}). The variables $x_n, h_{n+2}, h_{n+1} ,h_n$ in (\ref{AP:eq4.17})
are also invariants since the considered $\SL(2, \mathbb{R})$  group acts on the $y$ space only. We can call (\ref{AP:eq4.15}) a Schwarzian ODE. Then (\ref{AP:eq4.19}) is a
Schwarzian O$\Delta$S.
   Schwarzian derivatives play a prominent role in the theory of integrable systems
\cite{AP:ref15} and of dynamical systems \cite{AP:ref16, AP:ref17}.

\section{Numerical results} \label{AP:sec5}

\subsection{General procedure for testing the numerical schemes}
\label{AP:subsec5.1}
This section reports on the numerical experiments performed using the schemes described in the previous sections. The schemes are used to compute  the solution for initial value problems on a given interval. Before describing the results for each of the four classes of symmetries analyzed in this paper, we first describe some general procedures to implement and test the various methods.

\subsubsection{Reference solution}\label{AP:subsubsec5.1.1}
For test-problems for which an analytical solution is not available, a very accurate and reliable reference solution is computed numerically and used to assess the performance of the point symmetry preserving scheme. This is done using Matlab's standard adaptive Runge-Kutta scheme ODE45, with a very strict tolerance on the error set at $tol=10^{-9}$. The first step is to convert the $n$-th order Equation (\ref{AP:eq1.1}) for $y(x)$ into a system of $n$ first order ODEs for $u_1(x)=y(x)$, $u_2(x)=y'(x)$, \dots, $u_n(x)=y^{(n-1)}(x)$. Then Equation (\ref{AP:eq1.1}) becomes the system

\begin{equation}\label{AP:eq5.1}
u_1'=u_2 ,\quad  u_2'=u_3 ,\quad \dots ,\quad u_n'=F(x,u_1,u_2,\dots,u_n).
\end{equation}
Given initial conditions $u_1(x_0)$, $u_2(x_0), \dots, u_n(x_0)$, one then proceeds to compute the solution on the interval $[x_0,\, x_F]$, where the scheme adaptively selects the local integration step so that its local error estimates satisfies the imposed tolerance.
Those very high order, very accurate (and very costly numerically) solutions are used to generate start-up values as well as error estimations for the point symmetry preserving schemes as described next.

\subsubsection{Start-up values}\label{AP:subsubsec5.1.2}
The symmetry preserving schemes require a number of start-up values ($y_0=y(x_0)$, $y_1=y(x_1)$ for the second order case; also $y_2=y(x_2)$ for the third order cases). For given initial values $y(x_0)$, $y'(x_0)$, (and $y''(x_0)$ for the third order case), the start-up value $y_0=y(x_0)$ is directly available, while the values for $y_1$ and $y_2$ are obtained as the the numerical reference solution (obtained as described above) at the nodes $x_1$ and $x_2$.

\subsubsection{Error analysis}\label{AP:subsubsec5.1.3}
Given the discrete mesh $x_n$, $n=0,1,2, \dots ,N$  and corresponding solution $y_n$ generated by the point-symmetry preserving scheme, the corresponding errors are obtained by comparing $y_n$ with $y_{\rm{ref}}(x_n)$.
Although the user has no direct input on the actual mesh used by the Matlab's solver, it is possible for the user to request specific output points for the discrete solutions, so that given $x_n$, one can obtain a very reliable numerical approximation $y_{\rm{ref}}(x_n)$, accurate with the prescribed tolerance.

\subsubsection{Equivalent standard schemes}\label{AP:subsubsec5.1.4}
To better assess the new schemes proposed here,
their performance for various test-cases is compared with that of the standard finite difference schemes
that uses the same number of grid points $x_{s,n}$, $n=0,1,2, \dots, N$. Although the
point-symmetry preserving scheme finite mesh is typically non-uniform, for simplicity, the standard mesh is assumed to be, so that $x_{s,n}=x_0+n h$ with $h=(x_F-x_0)/N$.
The discrete standard scheme is obtained using the following standard
procedure (given here for the third order case, easily adapted for the second order
case). Given the four points $(x_{s,n-1},y_{s,n-1})$, $(x_{s,n},y_{s,n})$, $(x_{s,n+1}, y_{s,n+1})$, $(x_{s,n+2}, y_{s,n+2})$:
\begin{enumerate}
\item obtain the interpolating polynomial $P_3(x)$ through the four given points
\item evaluate analytically $P_3'(x_{s,n+1/2})$, $P_3''(x_{s,n+1/2})$, $P_3'''(x_{s,n+1/2})$, which gives:
\begin{eqnarray}
 P_3'(x_{s,n+1/2})
 &=\frac{1}{24 h}  (27(y_{s,n+1}-y_{s,n})-(y_{s,n+2}-y_{s,n-1}) )\label{eqnew1}\\ 
 P_3''(x_{s,n+1/2})
 &=\frac{1}{2 h^2}(y_{s,n+2}-(y_{s,n+1}+y_{s,n})+y_{s,n-1})\label{eqnew2}\\ 
 P_3'''(x_{s,n+1/2})
 &=\frac{1}{h^3}(y_{s,n+2}-3y_{s,n+1}+3y_{s,n}-y_{s,n-1})\label{eqnew3} 
\end{eqnarray}
\item substitute those expressions in the equation being discretized, evaluated at $x=x_{s,n+1/2}$.
\end{enumerate}

\subsection{\mathversion{bold}Numerical experiments for Example 1 (second order {\rm{ODE (\ref{AP:eq2.1})}})}\label{AP:subsec5.2}
Selecting $k=3$ in Equation (\ref{AP:eq2.1}) gives:
\begin{equation}\label{AP:eq5.2}
x^2 y''+4 x y' + 2y =(2 x y + x^2 y')^{1/2}
\end{equation}
to be solved for $x$ in the interval $[1,\, 3]$, with the initial conditions chosen as $y(1)=13/12$, $y'(1)=-1$.
The problem has an exact solution: $y_{\rm{ref}}(x)=x/12 + 1/x^2$.

The symmetry preserving scheme (\ref{AP:eq2.8}), (\ref{AP:eq2.9}) is used with the special choice $K=1$, so
that the mesh is uniform and the discrete scheme is given by:
\begin{equation}\label{AP:eq5.3}
x_{n+1}^2 y_{n+1}-2 x_n^2 y_n + x_{n-1}^2 y_{n-1}= ( {\textstyle \frac{1}{2}})^{1/2} h^{3/2} (x_{n+1}^2 y_{n+1}-x_{n-1}^2 y_{n-1})^{1/2}
\end{equation}
with $x_n=x_0 + n h$.  

The start-up values are given by $y_0=y(x=1)=13/12$ and
$y_1=y(x=1+h)=(1+h)/12+1/(1+h)^2$. The corresponding standard discrete scheme is given by:
\begin{eqnarray}\label{AP:eq5.4}
&(y_{s,n+1}-2 y_{s,n} + y_{s,n-1}) x_{s,n}^2 + 2 x_{s,n} h (y_{s,n+1}-y_{s,n-1})+ 2 h^2 y_{s,n} \\
&\qquad\qquad\qquad\qquad\qquad\qquad = h^2 \left(2 x_{s,n} y_{s,n} +x_{s,n}^2\frac{y_{s,n+1}-y_{s,n-1}}{2h}\right)^{1/2} \nonumber
\end{eqnarray}

Note that both the symmetry preserving scheme as well as the standard scheme
lead to a nonlinear problem to compute $y_{n+1}$, given $y_n$ and $y_{n-1}$. Those
algebraic nonlinear problems are solved using a standard fixed point iteration
until convergence.

Using the exact solution as reference, errors are computed for the numerical solutions using each of the two schemes, with mesh sizes $h=0.1$, $0.01$, $0.001$.
Those errors are reported in Table~\ref{AP:tab1}.

\begin{table}
\begin{center}
\caption{Discretization errors, Example~1\label{AP:tab1}}
\begin{tabular}{|c|c|c|c|}
\hline  Scheme & $h=0.1$ & $h=0.01$ & $h=0.001$\\ \hline
Sym.pres.& $6.04 \; 10^{-4}$&$7.26 \; 10^{-6}$&$7.39 \; 10^{-8}$\\ 
Standard & $4.72 \; 10^{-3}$&$7.54 \; 10^{-5}$&$7.86 \; 10^{-7}$\\ 
\hline
\end{tabular}
\end{center}
\end{table}

\begin{figure}[ht]
\includegraphics[width=\linewidth]{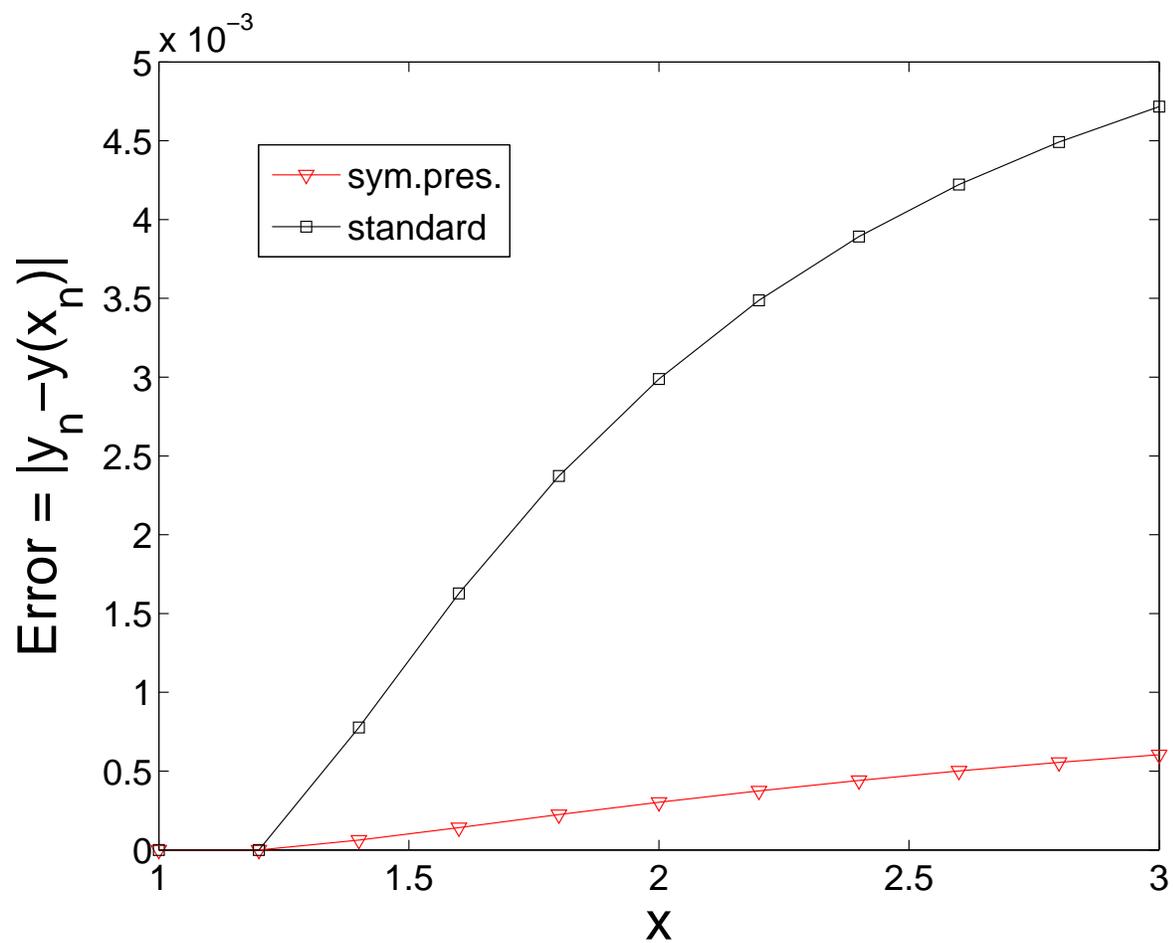}
\caption{
\label{AP:fig1}
Discretization errors for the symmetry preserving scheme and the standard scheme, Example 1.
}
\end{figure}

One observes from Table~\ref{AP:tab1} that both schemes are second order accurate,
as the error is
roughly divided by a factor $100$ whenever the mesh size is divided by $10$. Also,
the errors from the symmetry preserving schemes are smaller by a factor of
 $10$ compared to the errors obtained with the standard scheme with the same mesh size. This is achieved
without any additional computational cost, both schemes having the same
computational complexity. Figure~\ref{AP:fig1} shows the error as a function of $x$ for both schemes for mesh
size $h=0.1$. The gain from using the symmetry preserving scheme is obvious.

\subsection{Numerical experiments for Example 2 (third order {\rm ODE (\ref{AP:eq3.7})})}\label{AP:subsec5.3}
The test-case consists in solving Equation (\ref{AP:eq3.7}) for $K=1$, with $x$ in the interval $[0,\, 10]$ and with initial values $y(0)=0$, $y'(0)=-10$, $y''(0)=1$.
The lattice equation is chosen in the form (\ref{AP:eq3.15}), i.e.
\begin{equation}\label{AP:eq5.5}
\frac{\xi_1}{\xi_2}=\frac{\xi_2}{\xi_3}=\gamma.
\end{equation}

Start-up values for $x_0=0$, $x_1=h_0$, $x_2= 2 h_0$ are generated for a given $h_0$ using the Matlab solver, see Subsection~\ref{AP:subsec5.1}. The constant $\gamma$ in  (\ref{AP:eq5.5}) is then computed using the start-up points $(x_0,y_0)$, $(x_1,y_1)$, and $(x_2,y_2)$.

Given the three points $(x_{n-1},y_{n-1})$, $(x_n,y_n)$, $(x_{n+1},y_{n+1})$, the new point $(x_{n+2},y_{n+2})$ is obtained as the solution of the nonlinear system consisting of  Equations (\ref{AP:eq3.14}) (with $K =1$) and (\ref{AP:eq5.5}).
In the present set of experiments, the values $\alpha=\beta= {\textstyle\frac{1}{2}}$ were selected.
The resulting problem for $(x_{n+2},y_{n+2})$ is nonlinear, in particular the
mesh $x_n$ is non-uniform and completely coupled with the solution $y_n$.

The standard scheme is obtained by substituting the expressions in (\ref{eqnew1})(\ref{eqnew2})(\ref{eqnew3}) in  (\ref{AP:eq3.7}). 
It also leads to a nonlinear problem for $y_{n+2}$, but for that scheme, the mesh $x_n$ is assumed to be uniform and certainly completely decoupled from the solution $y_n$.

Table~\ref{AP:tab2} reports the numerical errors corresponding to various values for $h_0$ to start up the symmetry preserving schemes: $h_0=1$, $0.1$, $0.01$, which lead to respectively $14$, $130$ and $1297$ mesh nodes. The solutions with the standard schemes were computed on uniform meshes with the same number of nodes. 

\begin{table} 
\begin{center}
\caption{Discretization errors, Example 2 \label{AP:tab2}}
\begin{tabular}{|c|c|c|c|}
 \hline Scheme & $h=1 $& $h=0.1$ & $h=0.01$\\ 
  & $(N=14)$ & $(N=130)$ & $(N=1297)$\\ \hline
Sym.pres. &$ 2.14 \; 10^{-5}$&$2.98 \; 10^{-7}$&$6.45 \; 10^{-9}$\\ 
Standard&$ 4.20 \; 10^{-2}$&$5.83 \; 10^{-4}$&$6.01 \; 10^{-6}$\\ \hline
\end{tabular}
\end{center}
\end{table}

\begin{figure}[ht]
\includegraphics[width=\linewidth]{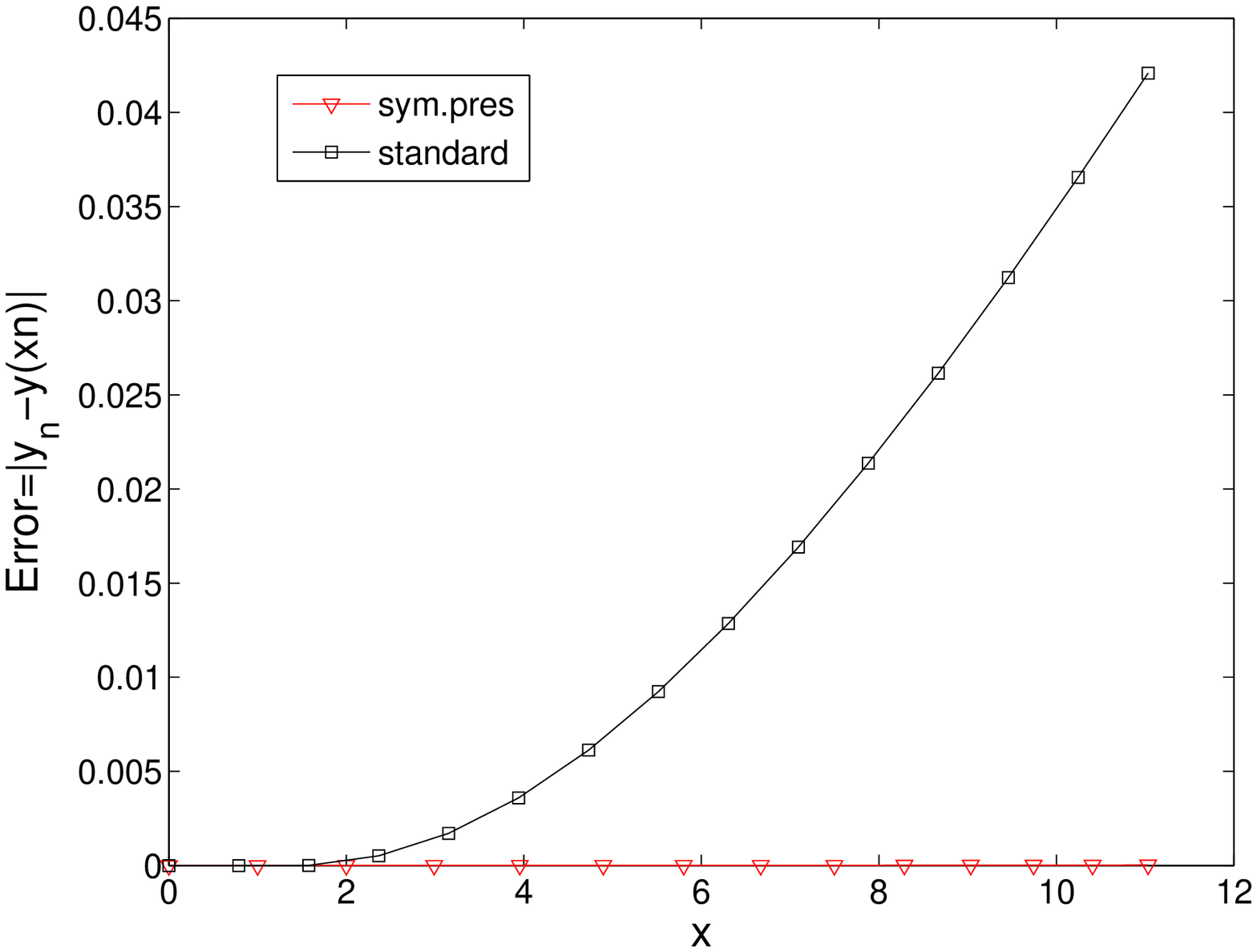}
\caption{
\label{AP:fig2}
Discretization errors for the symmetry preserving scheme and the standard scheme, Example 2, $h=1$.
}
\end{figure}

Both schemes appear to be effectively second order accurate, with the error in the symmetry preserving scheme smaller by a factor 1000. The discretization errors with both schemes are shown in Figure~\ref{AP:fig2}.

\subsection{Numerical experiments for Example 3 (third order {\rm ODE}  {\rm(\ref{AP:eq3.18})})} \label{AP:subsec5.4}

The test-case consists of solving Equation (\ref{AP:eq3.18}) for the special choice $F(I_1)= I_1^2$, which leads to the equation: 
\begin{equation}
(1+x^2)y'''+3 x y''=y''^2 (1+x^2)^{3/2}
\label{AP:eq5.6}
\end{equation}
The solution is sought for $x$ in the interval $[0 ,\, L]$, with $L$ to be selected below. The start-up values $\bigl(x_0=0, y_0=y(x_0)\bigr)$, $\bigl(x_1=h_0, y_1=y(x_1)\bigr)$, $\left(x_2= 2 h_0, y_2=y(x_2)\right)$ are obtained as before using an over-resolved numerical integrator.The procedure to generate the mesh $x_n$, $n=0, 1, 2, \dots, N$ and the corresponding discrete solution $y_n$ is as follows:
\begin{itemize}
\item {\bf Step 1.} Using the invariant Equation (\ref{AP:eq3.22}), one generates the complete mesh (for this particular case, it is independent of $y_n$). The constants in Equation (\ref{AP:eq3.22}) are taken as $a=1$, $b=-\gamma$, $c=0$ leading to
\begin{equation}\label{AP:eq5.5bis}
\frac{\xi_3}{\xi_4}=\frac{\xi_4}{\xi_5}=\gamma.
\end{equation}
The strategy to select $\gamma$ and compute the corresponding mesh is the same as the one used for Example 2, see discussion above.

\item {\bf Step 2.} Given the mesh $x_n$, solve the invariant Equation (\ref{AP:eq3.22}) for $y_{n+2}$ given $(x_{n-1},y_{n-1})$, $(x_n,y_n)$, $(x_{n+1},y_{n+1})$, and $x_{n+2}$.
\end{itemize}

Noting that $I_2=(1+x^2)^{(1/2)} dI_1/dx$, the equation being solved can be rewritten as 
$(1+x^2) dI_1/dx= I_1^2$.
The solution for $I_1(x)$ is therefore given by:
\begin{equation}\label{AP:eq5.7}
\frac{1}{I_1} = \frac{1}{I_{1,0}} - \arctan(x)
\end{equation}
where $I_{1,0}=I_1(x_0=0)$. This shows that $y''(x)$ will blow-up if $x=\tan(1/I_{1,0})$.
We assess the performance of the scheme for two cases, one with blow-up and one without.

\subsubsection{Blow-up case}\label{AP:subsubsec5.4.1}
The integration is performed for $x$ in the interval $[0 ,\, 11.2]$ with blow-up set up to occur at $x_b=11.25$. This is achieved by imposing $y''(0)=1/\arctan(x_b)$.
Three values for the initial $h_0$ are selected to be $h_0=0.1$, $0.01$, $0.001$, which lead to meshes with respectively $N=18$, $151$, $1484$ nodes.
Table~\ref{AP:tab3} reports the errors at $x_F=11.2$. Both the symmetry-preserving and the standard schemes appear to be of order $1$, with the errors from the symmetry-preserving schemes significantly smaller.

\begin{table}
\begin{center}
\caption{Discretization errors, Example~3, case with blow-up. \label{AP:tab3}}
\begin{tabular}{|c|c|c|c|}
\hline  Scheme & $h_0=0.1$ & $h_0=0.01 $ & $h_0=0.001 $\\ \hline
Sym.pres.&$ 8.82 \; 10^{-2}$&$9.88 \; 10^{-3}$&$3.93 \; 10^{-4}$\\ 
Standard &$ 7.03 \; 10^{-1}$&$1.10 \; 10^{-1}$&$1.68 \; 10^{-3}$\\ 
\hline
\end{tabular}
\end{center}
\end{table}

\begin{figure}[ht]
\includegraphics[width=\linewidth]{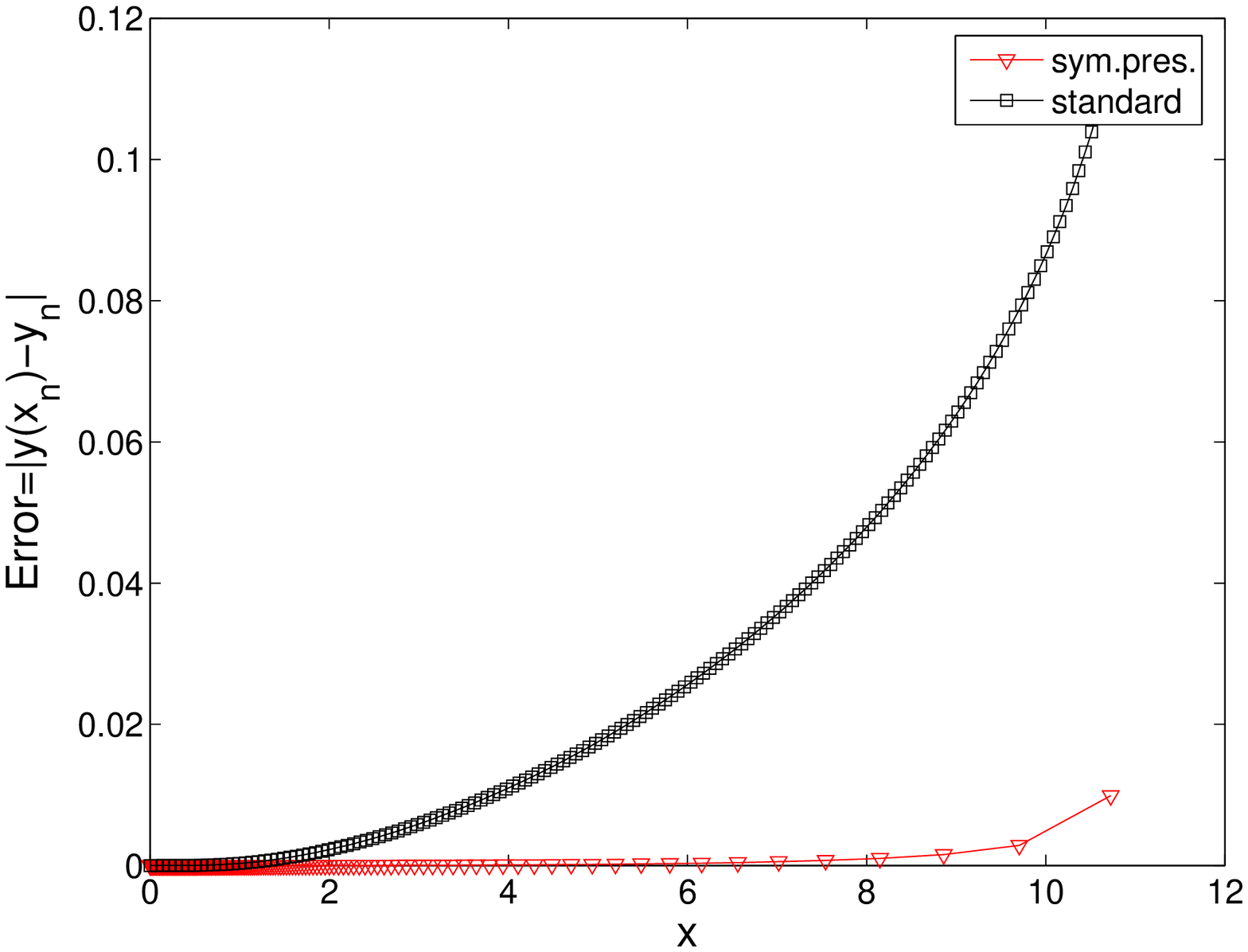}
\caption{
\label{AP:fig3}
Discretization errors for the symmetry preserving scheme and the standard scheme, Example 3, for the case with blow-up.
}
\end{figure}

Figure~\ref{AP:fig3} shows the behaviour of the discretization errors for both schemes, for the case $h_0=0.01$. 

\subsubsection{No blow-up case} \label{AP:subsubsec5.4.2}
This time, we select $y''(0)=-1/\arctan(x_b)$, so that blow-up will not occur for $x>0$. The numerical experiments are repeated with this new initial value. Table~\ref{AP:tab4} presents the errors at $x_F$ for various values of $h_0$, the conclusions are the same as for the blow-up case: both schemes appear to be first order accurate, with the symmetry-preserving scheme much more accurate.

\begin{table}
\begin{center}
\caption{Discretization errors, Example~3, case without blow-up. \label{AP:tab4}}
\begin{tabular}{|c|c|c|c|}
\hline  Scheme & $h_0=0.1$ & $h_0=0.01 $ & $h_0=0.001 $\\ \hline
Sym.pres.&$ 1.53 \; 10^{-3}$&$1.62 \; 10^{-5}$&$2.63 \; 10^{-6}$\\ 
Standard &$ 1.44 \; 10^{-1}$&$9.65 \; 10^{-3}$&$1.18 \; 10^{-4}$\\ 
\hline
\end{tabular}
\end{center}
\end{table}

\begin{figure}[ht]
\includegraphics[width=\linewidth]{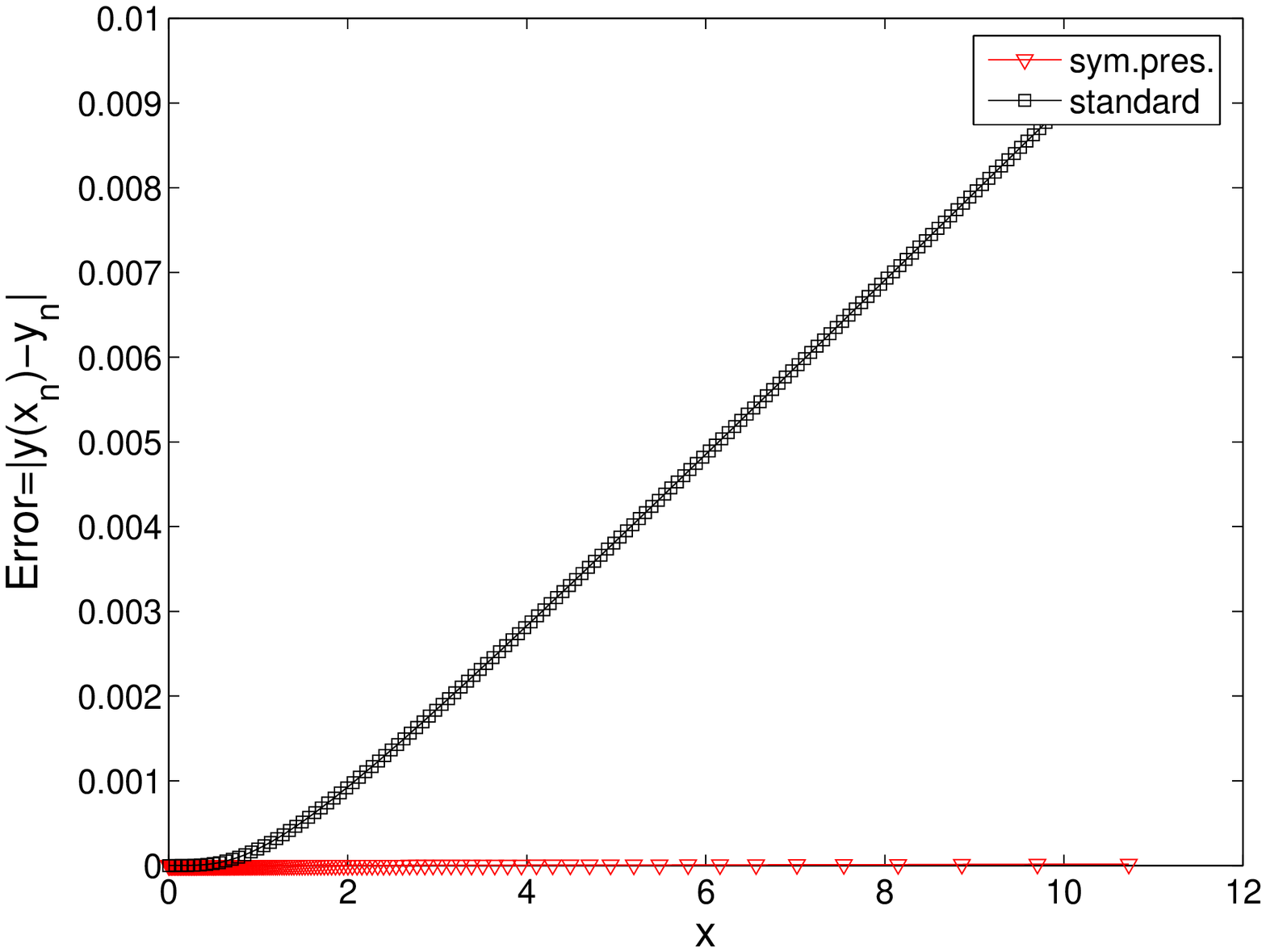}
\caption{
\label{AP:fig4}
Discretization errors for the symmetry preserving scheme and the standard scheme, Example 3, for the case without blow-up.
}
\end{figure}

Figure~\ref{AP:fig4} illustrates this behaviour for $h_0=0.01$.

\subsection{Numerical experiments for Example 4 (third order {\rm ODE (\ref{AP:eq4.5})})} \label{AP:subsec5.5}
The test-case consists of solving Equation (\ref{AP:eq4.5}) with $A=-1$, i.e.\ the difference equation
\begin{equation}
J_2=-J_1^{3/2}\label{AP:eq5.8}
\end{equation}
 on the lattice given by 
\begin{equation} \label{AP:eq5.9}
\frac{\xi_1}{\xi_2}=\gamma.
\end{equation}
with $J_2$, $J_1$ as in (\ref{AP:eq4.7}) and (\ref{AP:eq4.8}) and $\xi_i$ as in (\ref{AP:eq4.6}). The solution is sought for $x$ in the interval $[1, \, 16]$ with initial conditions $y(1)=0$, $y'(1)=0.1$, $y''(1)=0.1$. The start-up values $\bigl(x_0=1, y_0=y(x_0)\bigr)$, $\bigl(x_1=1+h_0, y_1=y(x_1)\bigr)$, $\bigl(x_2= 1+2 h_0, y_2=y(x_2)\bigr)$ are computed as in the other cases.
Given the three points $(x_{n-1},y_{n-1})$, $(x_n,y_n)$, $(x_{n+1},y_{n+1})$, the next point $(x_{n+2},y_{n+2})$ is obtained as the solution of the nonlinear system corresponding to the two symmetry preserving discrete Equations (\ref{AP:eq5.8}) and (\ref{AP:eq5.9}). The constant $\gamma$ in   (\ref{AP:eq5.9}) is computed based on the three start-up values. Table~\ref{AP:tab5} contains the errors with the symmetry preserving scheme and the standard scheme for this example, corresponding to various initial mesh sizes $h_0 = 0.2$, $0.01$, $0.005$.

\begin{table} 
\begin{center}
\caption{Discretization errors, Example~4 \label{AP:tab5}}
\begin{tabular}{|c|c|c|c|}
 \hline Scheme & $h=0.02 $& $h=0.01$ & $h=0.005$\\ 
  & $(N=149)$ & $(N=288)$ & $(N=567)$\\ \hline
Sym.pres.&$ 2.67 \; 10^{-4}$&$6.62 \; 10^{-5}$&$1.65 \; 10^{-5}$\\ 
Standard &$ 1.47 \; 10^{-3}$&$5.59 \; 10^{-4}$&$1.78 \; 10^{-4}$\\ \hline
\end{tabular}
\end{center}
\end{table}

\begin{figure}[ht]
\includegraphics[width=\linewidth]{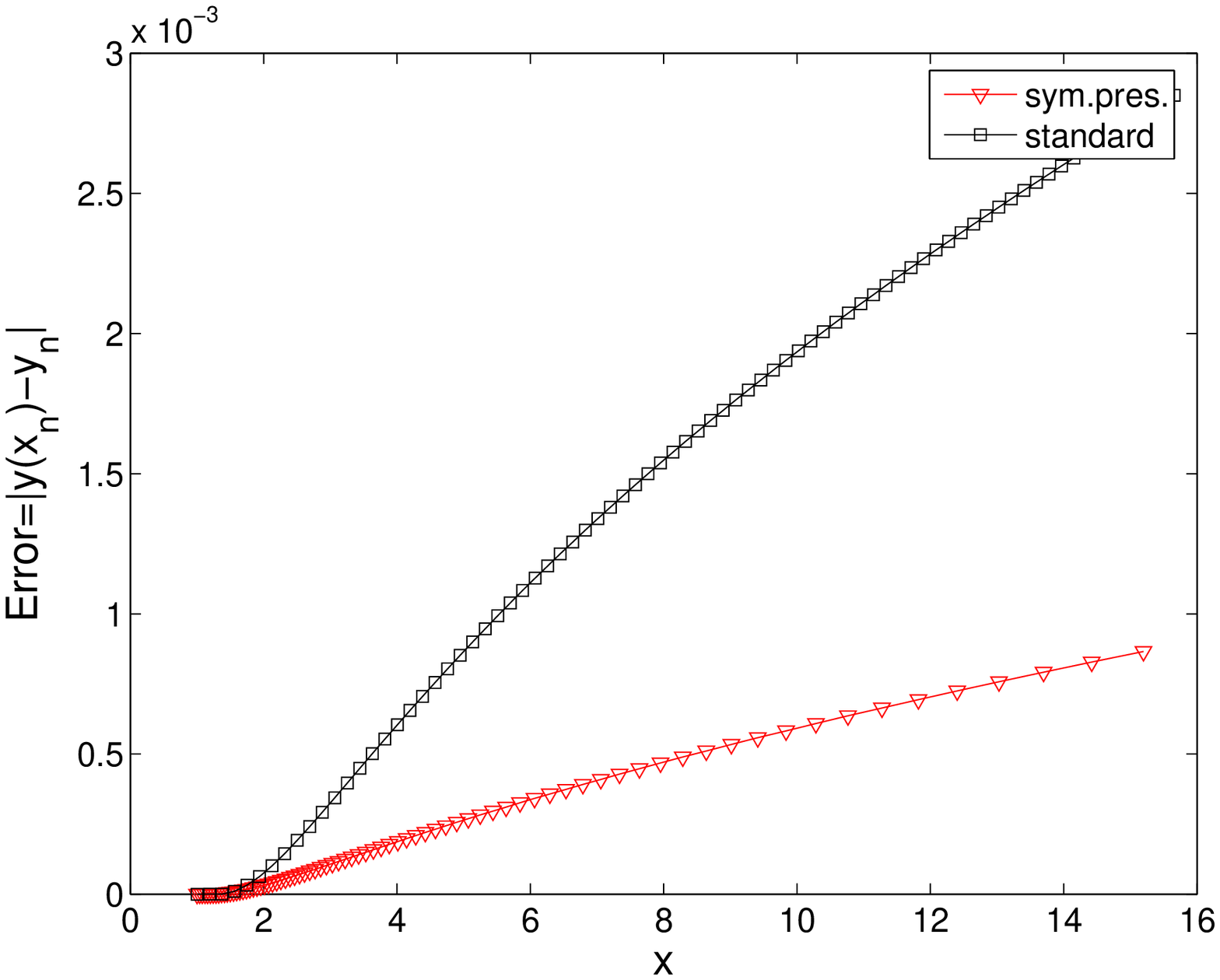}
\caption{
\label{AP:fig6}
Discretization errors for the symmetry preserving scheme and the standard scheme, Example 4.
}
\end{figure}

According to the results in Table~\ref{AP:tab5}, both schemes are second order accurate, with a much smaller error for the symmetry preserving scheme.
Figure \ref{AP:fig6} represents the discretization error behaviour for $h_0=0.04$. 

\subsection{Numerical experiments for Example 5 (third order {\rm ODE (\ref{AP:eq4.15})})} \label{AP:subsec5.6}

Numerical experiments are conducted with Equation (\ref{AP:eq4.15}) for the case $F(x)=sin(x)$:
\begin{equation} \label{AP:eq4.15bis}
\frac{1}{y^{'2}}\Biggl( y'y''' - \frac{3}{2} y^{''2}\Biggr) = sin(x).
\end{equation}

The solution is sought for $x$ in the interval $[0,\, 2]$ (also $[0,\, 6]$) with initial conditions $y(1)=0$, $y'(1)=-10$, $y''(1)=1$. A uniform mesh is used here, it is compatible with the difference invariants in (\ref{AP:eq4.19}). 
 With $h_n=h_{n+1}=h_{n+2}=h$ corresponding to the uniform mesh, the other invariant difference equation in (\ref{AP:eq4.19}) becomes:
\begin{equation}
R=4 ( 1 -\frac{h^2}{2} F(x_n,h))
\end{equation}
with $R$ defined in (\ref{AP:eq4.17}) as $R = (y_{n+2} - y_n)(y_{n+1} - y_{n-1})/((y_{n+2} - y_{n+1})(y_n - y_{n-1}))$ and where we select $F(x_n,h)=F(x_n+h/2)$ to achieve second order accuracy.
This leads to the following explicit expression for $y_{n+2}$:
\begin{equation} \label{pavel}
y_{n+2}=\frac{(y_{n+1}-y_{n-1})y_n-K (y_{n}-y_{n-1})y_{n+1}}{(y_{n+1}-y_{n-1})-K (y_{n}-y_{n-1})}
\end{equation}
where $K=4(1-\frac{h^2}{2} F(x_n+h/2))$. This explicit expression for $y_{n+2}$ is remarkably simple.

On the other hand, the standard scheme for the same problem is nonlinear. Substituting the finite difference approximations for $y', y''. y'''$ in (\ref{eqnew1}), (\ref{eqnew2}), (\ref{eqnew3}) in the ODE (\ref{AP:eq4.15bis}) leads to a nonlinear equation for $y_{n+2}$ to be solved iteratively.

\begin{figure}[h]
\includegraphics[width=\linewidth]{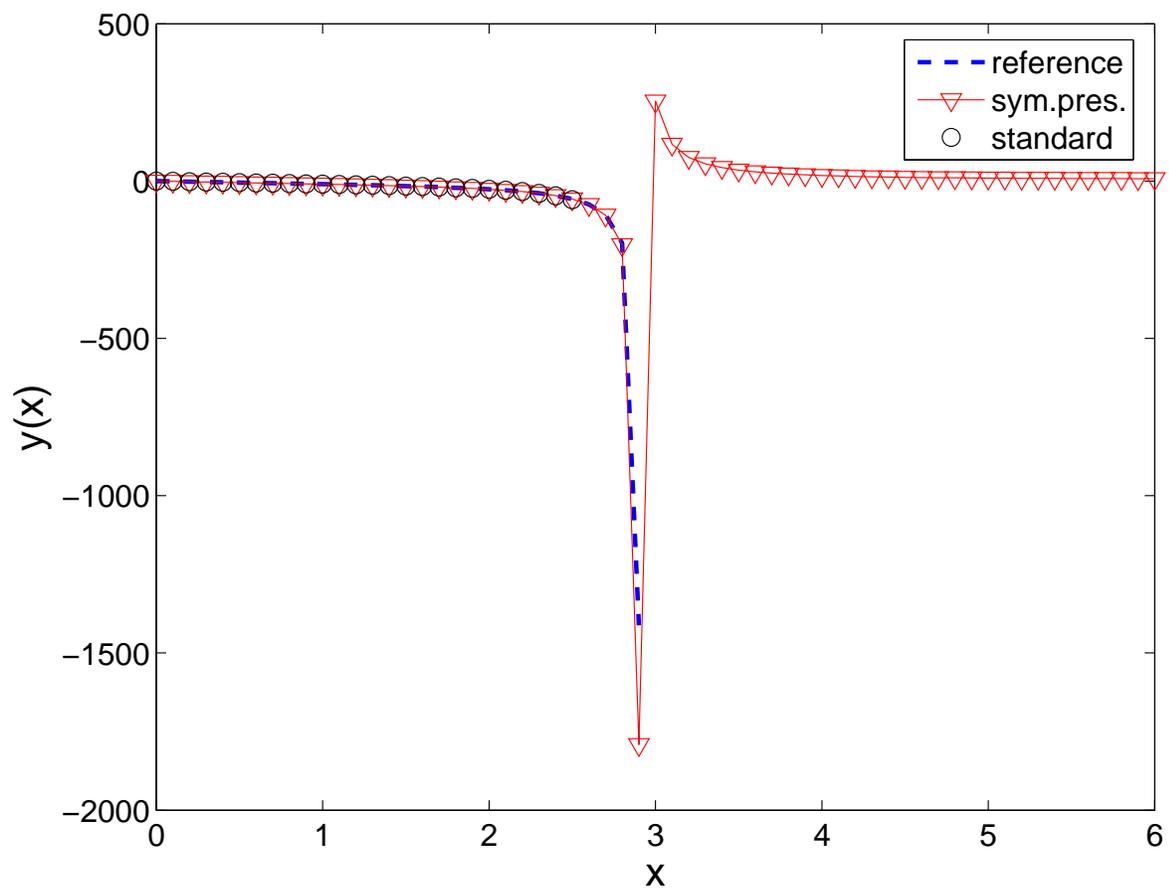}
\caption{
\label{AP:fig8}
Solution for the symmetry preserving scheme and the standard scheme, h=0.1, Example 5, on $[0,\, 6]$.
}
\end{figure}

\begin{figure}[h]
\includegraphics[width=\linewidth]{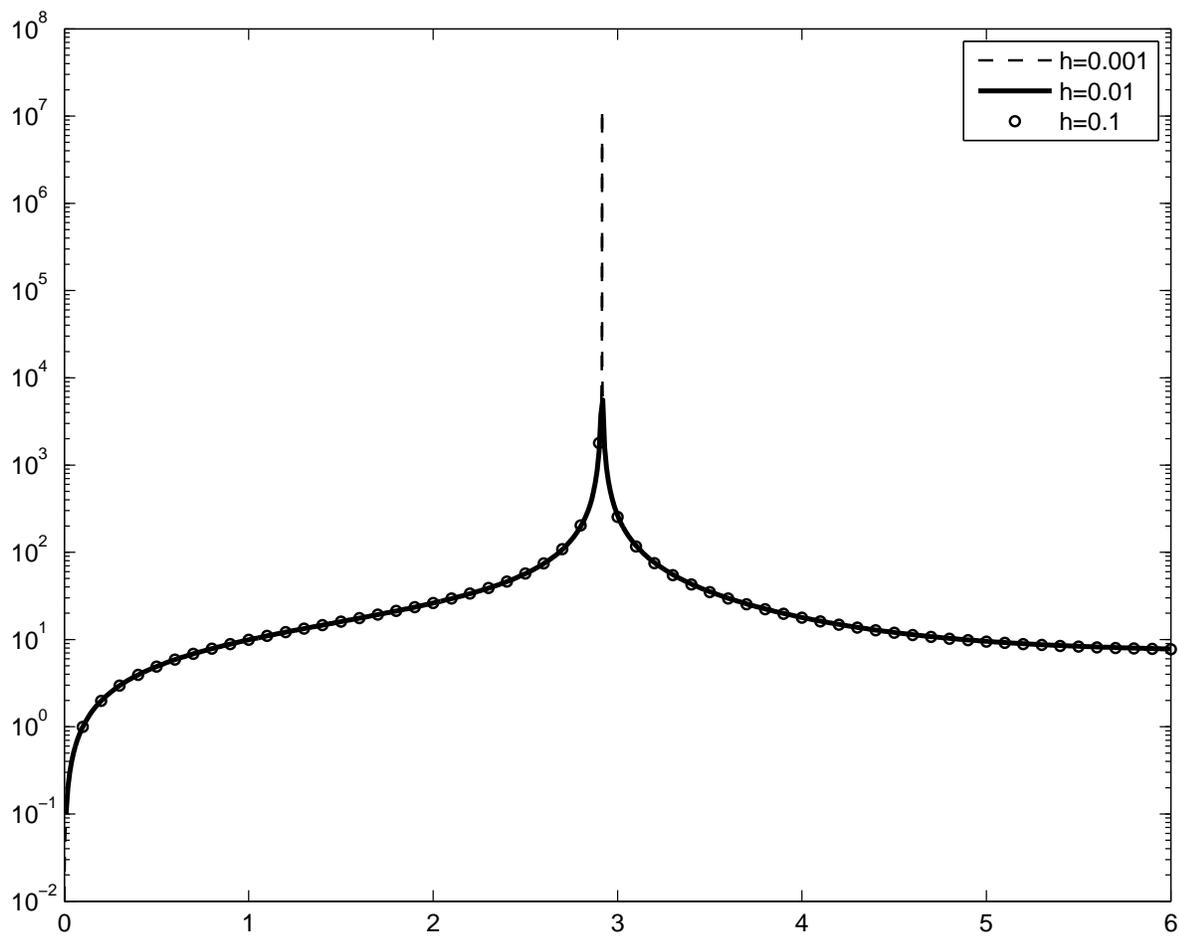}
\caption{
\label{AP:fig9}
Solution for the symmetry preserving scheme, h=0.1, 0.01, 0.001.  Example 5, on $[0,\, 6]$.
}
\end{figure}

First, we compare the discretization errors using the invariant scheme and the standard scheme on the interval $[0,\, 2]$ on which the solution is smooth. Table \ref{AP:tab6} reports those errors in terms of the mesh size $h$. Both schemes display a second order convergence rate. The standard scheme has errors which are smaller by a factor of 6, but in terms of computational efforts, the invariant scheme is much more efficient, as it gives an explicit formula for $y_{n+2}$ unlike the standard scheme that requires a nonlinear iterative solver at each step. 
However, if the integration interval is $[0,\, 6]$, remarkably different conclusions are obtained. The solution develops a singularity around $x=3$. At that point, both the standard scheme and the adaptive Runge-Kutta solver from Matlab fail to converge. On the other hand, the invariant scheme integrates right through the singularity. The solution obtained with the three schemes (reference, standard, invariant) is displayed in Figure \ref{AP:fig8} for the coarse resolution $h=0.1$. In Figure \ref{AP:fig9}, the solution is shown with the invariant scheme for three resolutions  $h=0.1,0.01, 0.001$. To better observe the behavior of the solution near the singularity, the plot uses a log scale, and the absolute value of the solution is shown. Excellent numerical convergence is observed, with the solutions corresponding to the three resolutions matching very closely each other (of course, the singularity is better captured by the finest mesh).

The most striking feature shown on Figure \ref{AP:fig8} and \ref{AP:fig9} is that the symmetry preserving difference scheme provides a numerical solution $u(x)$ for the entire region $0 \le x \le 6$ , $x \ne x_0$, even though the solution has a pole at $x_0$ close to $3$.
A similar phenomenon was observed in a previous study of a specific type of first order systems of ODEs, namely matrix Riccati equations \cite{AP:ref18, AP:ref19}. Matrix Riccati equations allow a ``nonlinear superposition formula" \cite{AP:ref19}, i.e. the general solution can be expressed algebraically in terms of a finite number of particular solutions.
The superposition formula is based on a nonlinear action of the group $\SL(N, \mathbb{R})$  
with $N=2$ for the Riccati equation itself. A numerical method based on this group theoretical superposition formula also made it possible to integrate around the poles of solutions \cite{AP:ref18} and to approach the poles from both sides.
A further relevant observation is that matrix Riccati equations can be discretized while preserving their superposition formulas \cite{AP:ref20, AP:ref21}. This discretization leads to fractional linear mappings similar in form to Equation (\ref{pavel})

\begin{table}[ht]
\begin{center}
\caption{Discretization errors, Example~5 \label{AP:tab6}}
\begin{tabular}{|c|c|c|c|}
 \hline Scheme & $h=0.1 $& $h=0.01$ & $h=0.001$\\ \hline
Sym.pres.&$ 3.10 \; 10^{-2}$&$3.13 \; 10^{-4}$&$2.96 \; 10^{-6}$\\
Standard &$ 5.07 \; 10^{-3}$&$5.01\; 10^{-5}$&$6.70\; 10^{-7}$\\ \hline
\end{tabular}
\end{center}
\end{table}

\section{Conclusions}
The basic motivation for this research program is that symmetries of a physical problem are an essential feature of the problem and should be incorporated in any mathematical model. In continuous descriptions, based on differential equations, this is taken for granted. In discrete descriptions, using difference equations, continuous symmetries are usually lost. It has been shown earlier \cite{AP:ref1, AP:ref2, AP:ref4, AP:ref5, AP:ref6, AP:ref7, AP:ref8a, AP:ref8b} that it is possible to construct difference schemes that possess the same symmetries as their continuous limits. To achieve this, it is necessary to use difference schemes (equations and meshes) constructed out of the invariants of the corresponding Lie groups.

In this article, we have considered second and third order ordinary differential equations with three, or four-dimensional symmetry groups. Our numerical experiments have shown that the accuracy of the symmetry preserving schemes is much better (sometimes three orders of magnitude better) than that of standard schemes at no significant additional cost. Example 5 has also shown that symmetry preserving schemes can also provide solutions when standard methods fail because of singularities.

Imposing that symmetries be preserved in a difference scheme usually still leaves some freedom in the scheme. For one, or two-dimensional symmetry groups standard schemes are very often among the symmetry preserving ones. Starting from dimension three this is usually not the case. In particular all examples treated in this article are such that standard schemes violate the symmetries.

We find the presented numerical experiments extremely encouraging. Future plans include an investigation of higher order ODEs and of systems of nonlinear ODEs from the point of view symmetry preserving discretizations. Also under study is the question of further optimizing the symmetry preserving schemes and further increasing their accuracy by exploiting the remaining freedom in the choice of lattices. The behaviour of solutions with singularities will be further studied. Finally, we are investigating the numerical implications of using symmetry preserving discretizations of partial differential equations \cite{AP:ref1, AP:ref11, AP:ref12, AP:ref13, AP:ref14}.

\subsection{Acknowledgments}
We thank Decio Levi for many very helpful discussions.
The research of A.~B. and P.~W. was partly supported by research grants from NSERC of Canada.

\section{List of captions}

\noindent
{\bf Figure 1}
Discretization errors for the symmetry preserving scheme and the standard scheme, Example 1.

\noindent
{\bf Figure 2}
Discretization errors for the symmetry preserving scheme and the standard scheme, Example 2.

\noindent
{\bf Figure 3}
Discretization errors for the symmetry preserving scheme and the standard scheme, Example 3, for the case with blow-up.

\noindent
{\bf Figure 4}
Discretization errors for the symmetry preserving scheme and the standard scheme, Example 3, for the case without blow-up.

\noindent
{\bf Figure 5}
Discretization errors for the symmetry preserving scheme and the standard scheme, Example 4.

\noindent
{\bf Figure 6}
Solution for the symmetry preserving scheme and the standard scheme, h=0.1, Example 5, on $[0,\, 6]$.

\noindent
{\bf Figure 7}
Solution for the symmetry preserving scheme, h=0.1, 0.01, 0.001.  Example 5, on $[0,\, 6]$.
  \newpage 
\section{Figures}

\vfill
\newpage

\vfill
\newpage

\vfill
\newpage

\vfill
\newpage

\vfill
\newpage

\vfill
\newpage
\end{document}